# Tunneling time in the case of frustrated total reflection

D. MUGNAI (*), A. RANFAGNI (*), L. RONCHI (**)

ABSTRACT- *The tunneling time is here investigated by means of an electromagnetic model, for a system where a gap, between two parallel planes, acts as a classically-forbidden region for an impinging pulse with incidence angle larger than the critical angle. In all cases of frustrated total reflection we obtain a superluminal behavior both for phase and group delays.*

## 1. Introduction

The tunneling process represents one of the most non-classical predictions of quantum mechanics. Inside this process, one of the most debated question is the "traversal time" that a system spends during its passage through a barrier. The question "how much time does tunneling take?" is certainly not new, but it has not yet been resolved due also to the difficulty in obtaining experimental confirmation. The several models hitherto proposed for evaluating the traversal time can be roughly grouped into two main trends: the semiclassical model and the phase time model. A third model has been proposed by Büttiker and Landauer: it can be considered as a sort of link between the two above, since it tends to reproduce the semiclassical results for energy well below the top of the barrier and the phase-time model above the barrier (1-3).

In the semiclassical model time $t_s$ is defined as the ratio between space $d$ and the classical velocity $v = \hbar|k|/m$, where $k = [2m(E-V_0)]^{1/2}/\hbar$ is the momentum and $m$ is the mass of the particle. It shows a singularity

(*) Istituto di Ricerca sulle Onde Elettromagnetiche "Nello Carrara" - CNR, Via Panciatichi 64, 50127 Firenze, Italy.
(**) Fondazione Giorgio Ronchi, Via San Felice a Ema 20, 50125 Firenze, Italy.

when the energy $E=\hbar\omega_0$ of the incident particle is equal to the barrier height $V_0$. Since in the classically-forbidden region, for $E<V_0$, the momentum is an imaginary quantity, the semiclassical time is also imaginary.

The absolute value of $t_s$ is given by

$$[1] \qquad t_s = d\sqrt{\frac{m}{2\,|E-V_0|}}.$$

In the phase-time model, the complex transmission coefficient of the barrier $T^{1/2}\exp(i\Delta\phi)$ is considered, and the tunneling time $t_\phi$ is expressed as the derivative of the phase shift $\Delta\phi$ with respect to energy $E$, that is,

$$[2] \qquad t_\phi = \hbar\frac{\partial(\Delta\phi)}{\partial E}.$$

For a rectangular potential barrier, the resulting expression for $E<V_0$ is (4)

$$[3] \qquad t_\phi = \frac{m}{\hbar k_1 \kappa}\left[\frac{\varepsilon^4 \sinh(2d\kappa) - 2dk_1^2\kappa(k_1^2 - \kappa^2)}{4k_1^2\kappa^2 \cosh^2(d\kappa) + (k_1^2 - \kappa^2)^2 \sinh^2(d\kappa)}\right],$$

where $\varepsilon=(2mV_0)^{1/2}/\hbar$, $k_1=(2mE)^{1/2}/\hbar$ and $\kappa=(\varepsilon^2-k_1^2)^{1/2}$. For $E>V_0$, we have to replace, in Eq. [3], $\kappa$ with $ik$.

Lastly, in the Büttiker-Landauer model, the traversal time $t_{BL}$ is defined as

$$[4] \qquad t_{BL} = \left|i\hbar\frac{\partial}{\partial E}\left(\ln T^{1/2} + i\Delta\phi\right)\right|.$$

For a rectangular potential barrier, for $E < V_0$, the real $t_R$ and imaginary $t_I$ parts of the complex quantity appearing in Eq. [4] are (5)

$$[5a] \qquad t_R = \frac{mk_1}{\hbar\kappa}\left[\frac{2\kappa d(\kappa^2 - k_1^2) + \varepsilon^2 \sinh(2\kappa d)}{4k_1^2\kappa^2 + \varepsilon^4 \sinh^2(\kappa d)}\right]$$

$$[5b] \qquad t_I = \frac{m\varepsilon^2}{\hbar\kappa^2}\left[\frac{(\kappa^2 - k_1^2)\sinh^2(\kappa d) + (\varepsilon^2\kappa d)\sinh(2\kappa d)}{4k_1^2\kappa^2 + \varepsilon^4 \sinh^2(\kappa d)}\right].$$

Also in this case, when $E > V_0$, we have to replace $\kappa$ with $ik$. In Fig. 1, we show the three models mentioned above translated into the electromagnetic framework that is by taking $\varepsilon = 2\pi\nu_0/c$, $k_1 = 2\pi\nu/c$, and once the substitution $m/\hbar \to 2\pi\nu/c^2$ is made in the prefactor of Eqs. [3] and [5]. In this framework, the semiclassical time of Eq. [1] becomes $t_s = 2\pi\nu d/c^2\kappa$.

Tunneling also occurs in optics, making it possible to establish a tight analogy between particle motion and electromagnetic wave propagation. The implications of this topic, which are ultimately connected to the question on particle-wave dualism, have recently been discussed also in connection with relativistic problems, for the possibility of observing superluminal motions (that is, with velocity greather than light speed in vacuum). In other words, the question "Is quantum tunneling faster than light?" is now brought to the bench (6, 7).

As is known, considerable difficulties are encountered when performing a direct measurement of tunneling time because of the very short times involved, which are approximately given by the ratio between some wavelength of the wavefunction and a velocity comparable with that of the light. With a solid-state device, the tunneling time involved may be, typically, of the order of femtoseconds. For optical tunneling in the visible region, the magnitude of the tunneling time is of the order of picoseconds or less; however a decisive increase of this time has been obtained by increasing the wavelength up to microwaves. In this way, the time scale is magnified up to nanoseconds, and the measurements can be easily performed (8-10). An experimental device suitable for simulating quantum tunneling consists of an undersized waveguide in which evanescent modes can take place. In analogy with the quantum tunneling of a particle through a barrier, the waveguide can be regarded as a one-dimensional barrier for electromagnetic waves. The analogy, however, goes beyond what is outlined, since quantum tunneling and evanescent waves are described by closely-related wave equations (11). The fact that the results of such a simulation are best described by a quantum-mechanical model, suitably

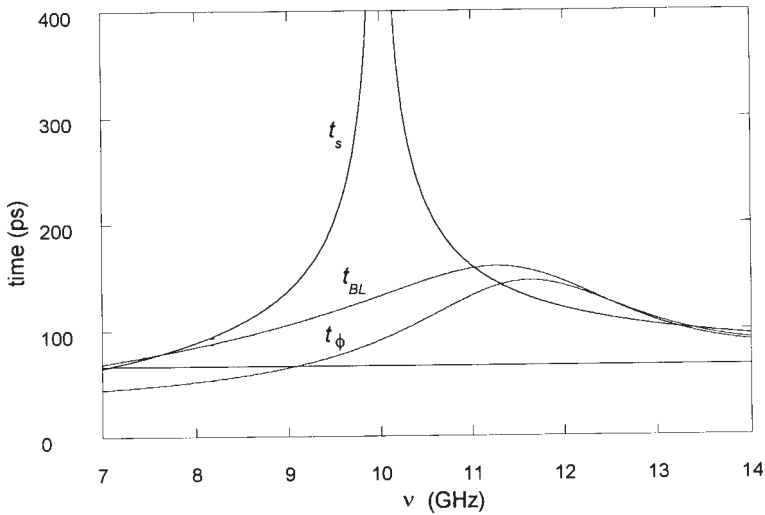

Fig. 1

Quantum mechanical tunneling time models, for a rectangular potential barrier, translated into the electromagnetic framework, as a function of the frequency. The lines denoted by $t_s$, $t_{BL}$, and $t_\phi$ represent the semiclassical, Büttiker-Landauer and phase-time models, respectively, calculated for a gap of 2 cm and a cutoff frequency of 10 GHz. The straight line represents the luminal ($v=c$) limit: the $t_\phi$ curve becomes superluminal below about 9 GHz while $t_{BL}$ and $t_s$ cross this limit below about 7 GHz.

translated into the classical electromagnetic framework, can be considered as proof that quantum tunneling can actually be simulated by these kinds of experiments. The dealy predicted by such a procedure can, within certain limits, be paradoxically small, implying a barrier-traversal velocity greater than the speed of light $c$ (Hartman effect (12)).

The delay time in an optical tunneling experiment was also measured using a two-photon interference method. The peak of the coincidence profile (the width of which is of the order of 100 fs) in the presence of the barrier is found "one" femtosecond *earlier* with respect to photons traveling in the vacuum at the light speed $c$. Although the apparent tunneling velocity (~1.7$c$) is superluminal, this was not considered to be a genuine signal velocity and the Einstein causality not violated (13).

Recently, delay time measurements in a diffraction experiment with microwaves have been performed, both for the phase delay (obtained directly) and for the group delay (deduced from the phase-dealy results) in the range of a few to hundreds of ps. The experimental setup consisted of a grating followed by a paraffin prism. For wavelengths greater than the period of the grating (3 cm), the diffracted waves (of non zero order) were evanescent waves which were transformed in real waves by the paraffin prism. The results obtained demonstrated that superluminal behaviour of the phase and group velocities was attained (14).

In the present paper, we theoretically approach (in Sec. 2) the case of evanescent waves which arise after total reflection from a paraffin prism (instead of a grating). The results of this analysis are then discussed in Sec. 3.

## 2. Theoretical analysis

We consider here a very simple system formed by two half-spaces, limited by the plane-parallel boundaries $\Pi_1$, $\Pi_2$, at $z=0$ and $z=d$, respectively, filled with a homogeneous and non-dispersive medium of refractive index $n$ (= 1.49 for the paraffin), and separated by a vacuum gap (as shown in Fig. 2). With this type of schematization, we cannot take into account the multiple reflections which take place in the experimental apparatus.

A plane wave e.m. pulse, $\mathbf{E}^i$, $\mathbf{H}^i$, impinges on the gap from the left, with an incidence angle larger than the limit angle $i_0=\sin^{-1}(1/n)$ ($\simeq 42°$ for the paraffin). Using standard methods, we evaluate the plane wave transmitted pulse $\mathbf{E}^t$, $\mathbf{H}^t$, for $z > d$ and its deformation with respect to the impinging pulse. From the comparison among them, it is possible to evaluate the time taken by the pulse to "travel" through the gap.

A Cartesian reference system $\mathbf{i}$, $\mathbf{j}$, $\mathbf{k}$ (cordinates x, y, z) is chosen with the origin $O$ on the plane $\Pi_1$, $\mathbf{k}$ normal to the planes $\Pi_1$, $\Pi_2$, and the x-axis $\mathbf{i}$ in the same plane as $\mathbf{k}$ and the direction of propagation $\mathbf{s}^i$ of the impinging field. For $\mathbf{s}$, we can therefore write

[6] $$\mathbf{s}^i = \alpha\mathbf{i}+\gamma\mathbf{k}, \text{ with } \gamma>0$$

For the sake of simplicity, we consider only the TE case, in which the impinging wave satisfies the condition $\mathbf{E}^i\cdot\mathbf{k}=0$ (and $\mathbf{E}^i\cdot\mathbf{s}^i=0$). Accordingly, we can write

[7] $$\mathbf{E}^i=E_y^i\,\mathbf{j},\quad \mathbf{H}^i = (1/Z)\mathbf{s}^i\wedge\mathbf{E}^i$$

where $Z=Z_0/n$, $Z_0$ being the free space impedance.

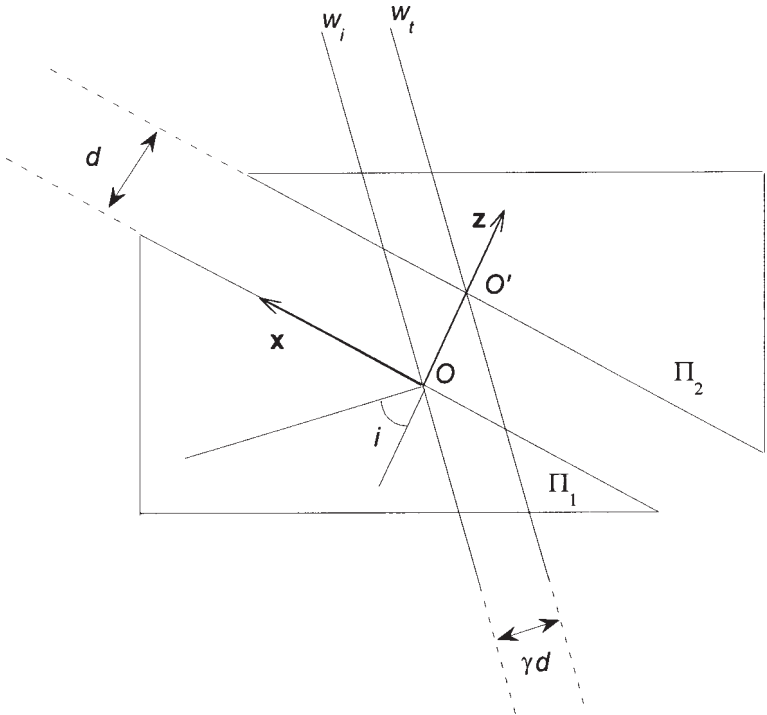

Fig. 2

Two paraffin-prisms in the total reflection condition with the coordinate system adopted in the theoretical analysis.

It is well known that boundary conditions (namely the continuity conditions of the tangential components of the fields at the surfaces of discontinuity of the refractive index imply that:

– in the half-space $z<0$, the field is the superposition of two waves, the impinging wave $\mathbf{E}^i$, $\mathbf{H}^i$ and a reflected wave $\mathbf{E}^r$, $\mathbf{H}^r$, (with $\mathbf{H}^r=(1/Z)\mathbf{s}^r\wedge\mathbf{E}^r$) propagating in the direction specified by the unit vector $\mathbf{s}^r=\alpha\mathbf{i}-\gamma\mathbf{k}$;

- in the gap, the field is the superposition of two evanescent waves, one attenuated in the positive direction of **k** and the other attenuated in the reverse direction –**k**. The former will be denoted as "progressive" wave **E**$^+$, **H**$^+$, with **H**$^+$=(1/$Z_0$)**s**$^+$∧**E**$^+$, the latter as "regressive" wave **E**$^-$, **H**$^-$, with **H**$^-$=(1/$Z_0$)**s**$^-$∧**E**$^-$. For **s**$^+$ and **s**$^-$, which turn out to be complex vectors of unit modulus, we have **s**$^+$=$n\alpha$**i**+$i\Gamma$**k**, **s**$^-$=$n\alpha$**i**–$i\Gamma$**k**, with $\Gamma=(n^2\alpha^2-1)^{1/2}=(n^2-1-n^2\gamma^2)^{1/2}>0$;

- in the half-space $z>d$, the field is composed of a single transmitted wave, **E**$^t$, **H**$^t$, with **H**$^t$=(1/$Z$)**s**$^i$∧**E**$^t$, propagating in the same direction **s**$^i$ as the impinging wave.

Naturally, the continuity conditions also make it possible to determine the amplitudes of the reflected and transmitted fields.

*A. Monochromatic case*

First, let us consider a plane monochromatic impinging wave, namely

[8] $$E_y^i = E_0 \exp\left[i\frac{\omega}{c}n(\alpha x+\gamma z)\right]\exp(-i\omega t)$$

and write the reflected, transmitted, progressive and regressive waves in the form

[9a] $$E_y^r = E_0\rho\exp\left[i\frac{\omega}{c}n(\alpha x-\gamma z)\right]\exp(-i\omega t)$$

[9b] $$E_y^t = E_0\tau\exp\left[i\frac{\omega}{c}n(\alpha x+\gamma(z-d))\right]\exp(-i\omega t)$$

[9c] $$E_y^+ = E_0 p\exp\left[i\frac{\omega}{c}(n\alpha x+i\Gamma z)\right]\exp(-i\omega t)$$

[9d] $$E_y^- = E_0 r\exp\left[i\frac{\omega}{c}(n\alpha x-i\Gamma z)\right]\exp(-i\omega t)$$

where ρ denotes the amplitude reflection coefficient, more precisely, the (complex) amplitude of the reflected wave at *O*, τ the transmission coefficient of the gap, more precisely, the amplitude of the transmitted field and the regressive waves at *O*, respectively. From the continuity conditions for the tangential component of the electric fields across $\Pi_1$ and $\Pi_2$, we obtain

[10] $$\begin{aligned}1+\rho&=p+r \quad \text{at } z=0\\ pe_1+re_2&=\tau \quad \text{at } z=d,\end{aligned}$$

where $e_1=\exp[-(\omega/c)\Gamma d]$, $e_2=1/e_1$. Analogously, from the continuity conditions for the tangential components of the magnetic fields at $z=0$ and $z=d$, we obtain

[11]
$$in\gamma(1-\rho)=-\Gamma(p-r)$$
$$in\gamma\tau=-\Gamma(pe_1-re_2).$$

By solving the linear system formed by Eqs. [10] and [11], we easily obtain

[12] $\quad p=\dfrac{e_2}{2\Gamma}(\Gamma-in\gamma)\tau, \quad r=\dfrac{e_1}{2\Gamma}(\Gamma+in\gamma)\tau, \quad \rho=\dfrac{\tau}{2\Gamma}\left[\Gamma(e_1+e_2)+in\gamma(e_1-e_2)\right]-1,$

and

[13]
$$\tau=\tau(\omega)=i\frac{4n\gamma\Gamma}{\Delta_1^2}\frac{\exp\left[-\omega\Gamma(d/c)\right]}{1-\exp\left[-2\omega\Gamma(d/c)-4i\phi\right]}$$
$$=iK\frac{1}{e^{\alpha\omega+2i\phi}-e^{-\alpha\omega-2i\phi}},$$

where

[14] $\quad K=\dfrac{4n\gamma\Gamma}{n^2\gamma^2+\Gamma^2}, \quad a=\dfrac{\Gamma d}{c}, \quad \phi=\arctan\left(\dfrac{\Gamma}{n\gamma}\right), \quad 0<\phi<\pi/2.$

It appears from Eq. [13] that the gap behaves like a low-pass filter, since for $|\omega|\to\infty$, $\tau\to 0$. Figure 3 shows $\tau(\omega)$ plotted versus $a\omega$, and its argument for two values of $\gamma$. We note that the absolute value of $\tau(\omega)$ is slowly varying with $\gamma$, while its argument substantially changes when $\gamma$ varies. More precisely, if $\phi>\pi/4$ ($\Gamma>n\gamma$), we have $arg(\tau)$ passing from positive (for negative $\omega$) to negative values (for positive $\omega$). The behavior of $arg(\tau)$ is just the opposite for $\phi<\pi/4$, and is equal to 0 everywhere for $\phi=\pi/4$ ($\Gamma=n\gamma$): in this case, $\tau(\omega)$ is real.

The expression $\tau(\omega)$ allows us to evaluate the traversal time of the gap. We consider the impinging wavefront $w_i$ through $O$ (at $x=0$, $z=0$) and the transmitted wavefront $w_t$ passing through $O'(x=0$ and $z=d)$ as in Fig. 2. According to Eqs. [8] and [9b], the phase $\phi_i$ of the former is $\phi_i=-\omega t$; that of the latter is $\phi_t=arg(\tau)-\omega t$. It follows that a given value of the phase, say 0, is at $O$ at time $t=0$, and at $O'$ at time $arg(\tau)/\omega$. The conclusion is that the phase takes as a time in passing from $O$ to $O'$, that is, to travel distance $\gamma d$. Note that for $\omega=0$, $t_{ph}=\arctan[(n^2\gamma^2-\Gamma^2)d/2n\gamma c]$.

[15] $\quad\quad\quad\quad\quad t_{ph}=arg(\tau)/\omega$

As for the group delay, we have (14)

$$t_{gr}=t_{ph}+v\frac{\partial t_{ph}}{\partial v}=t_{ph}+\omega\frac{\partial}{\partial\omega}\left[\frac{arg(\tau(\omega))}{\omega}\right].$$

From Eq. [13] we have

[16] $$arg(\tau(\omega)) = \arctan\left[\frac{n^2\gamma^2 - \Gamma^2}{2n\Gamma\gamma}\tanh\left(\frac{\omega\Gamma d}{c}\right)\right],$$

hence $t_{gr}$ can be obtained directly as

[17] $$t_{gr} = \frac{\partial}{\partial\omega}arg(\tau(\omega)) = \frac{2n\Gamma\gamma a(n^2\gamma^2 - \Gamma^2)}{(2n\Gamma\gamma)^2\cosh^2(a\omega) + (n^2\gamma^2 - \Gamma^2)^2\sinh^2(a\omega)}.$$

In Fig. 4*a*, *b*, *c* we show results of $t_{ph}$ as a function of *d*, $\omega$ and $\Gamma$, respectively together with time $t_l$ corresponding to real waves traveling distance $\gamma d$ at velocity *c*. Results of $t_{gr}$, as a function of $\Gamma$, are reported in Fig. 5 for some values of $\nu$ ($\omega=2\pi\nu$). The results obtained show a clear superluminal behavior, both in the phase and group delays, which is strongly dependent on the frequency, as well as on $\Gamma$: in particular for $\Gamma=n\gamma$, i.e. for $\gamma^2=(n^2-1)/2n^2$, the phase and group delays go to zero. For higher values of $\Gamma$ they became negative.

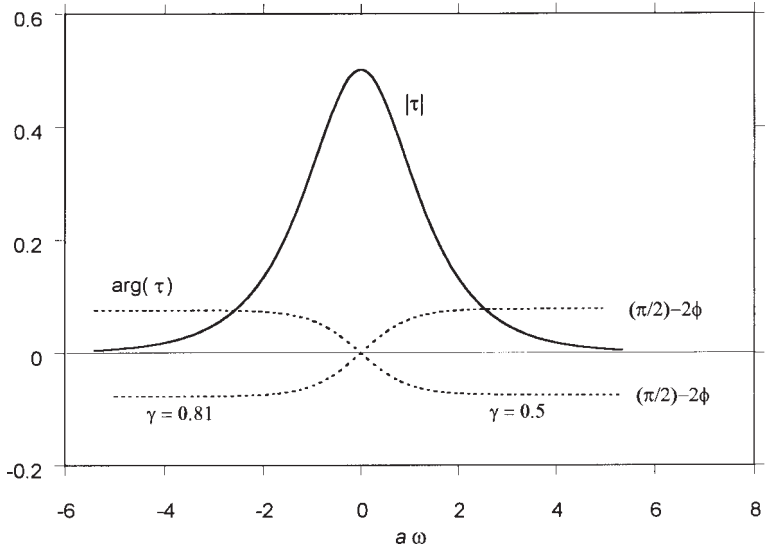

FIG. 3

Absolute value and argument of $\tau(\omega)$ as given by Eq. [13] for $d=2$ cm, $n=1.49$ and two values of $\gamma$.

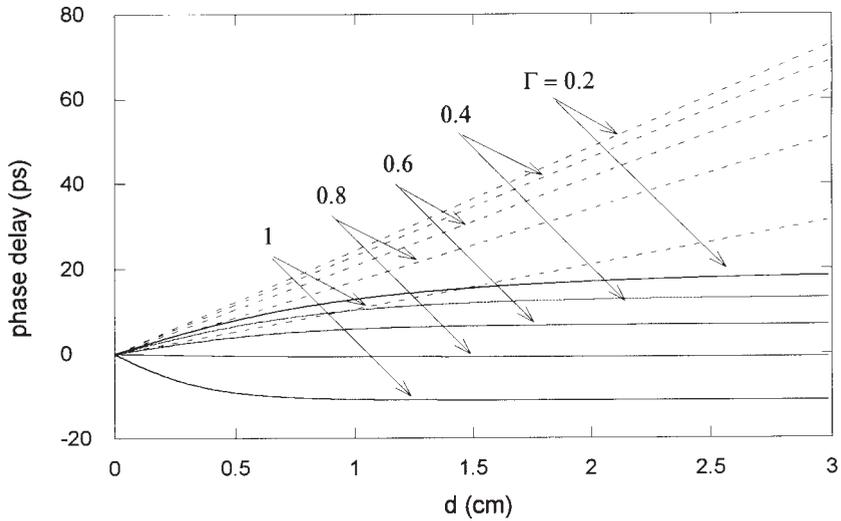

FIG. 4a

Phase delay (continuous line) between the wavefronts $w_i$ and $w_t$ (Fig. 2), as deduced from Eq. [15] as a function of the gap $d$ for some values of the parameter $\Gamma$; dashed lines represent the delay of a real wave traveling at velocity $c$ for the same set of parameter values.

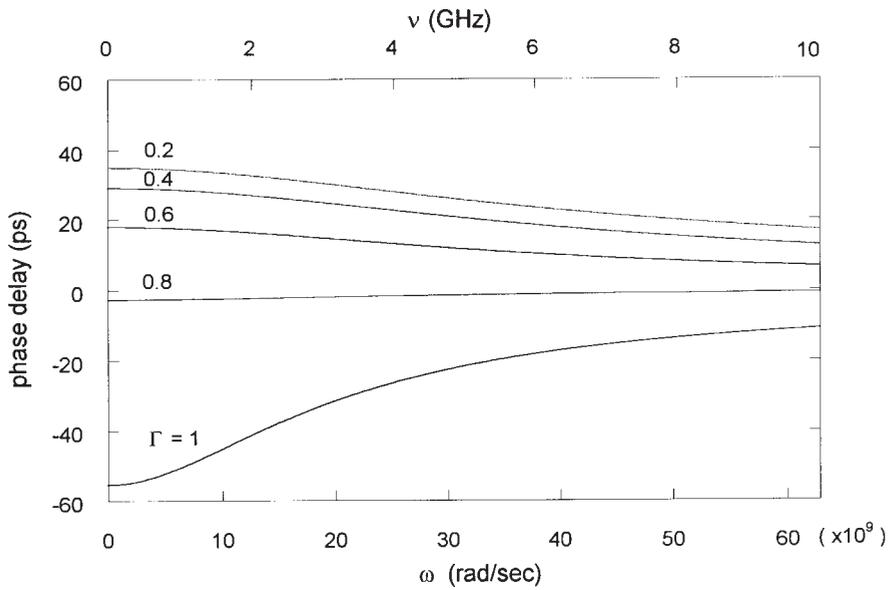

FIG. 4b

Phase delay, as a function of $\omega$, for some values of $\Gamma$ as in Fig. 4a.

## B. Case of an impinging spectrum

Equation [13] holds for a monochromatic plane wave. If the impinging wave has the character of a temporal pulse, Eq. [8] has to be replaced by

$$[18] \quad E_y^i = \frac{E_0}{2\pi} \int_{-\infty}^{\infty} A(\omega) \exp\left[i\frac{\omega}{c} n(\alpha x + \gamma z)\right] \exp(-i\omega t) \, d\omega,$$

where $A(\omega)$ is the incident spectrum, and the transmitted field can be witten as

$$[19] \quad E_y^t = \frac{E_0}{2\pi} \int_{-\infty}^{\infty} A(\omega)\tau(\omega) \exp\left[i\frac{\omega}{c} n\left[\alpha x + \gamma(z-d)\right]\right] \exp(-i\omega t) \, d\omega,$$

with $\tau(\omega)$ given by Eq. [13].

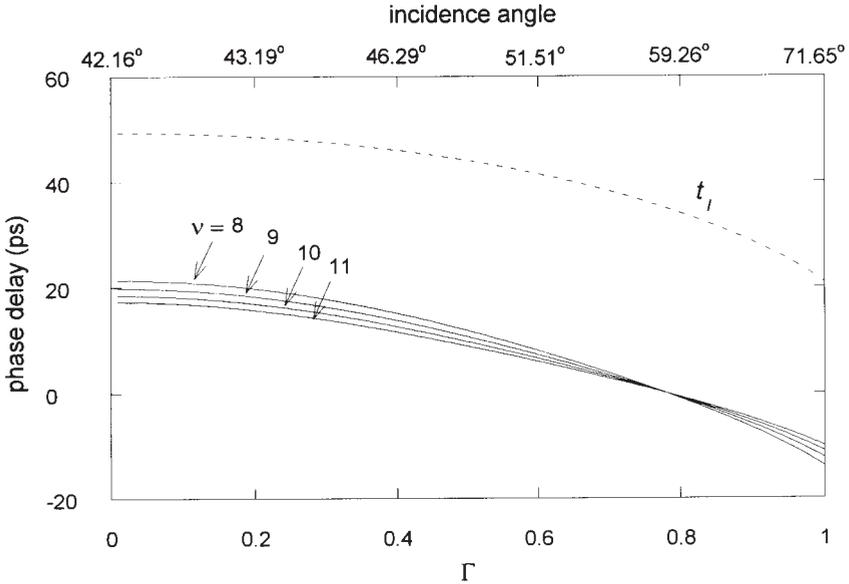

FIG. 4c

Phase delay (continuous lines) as a function of $\Gamma$ (or of incidence angle) for some value of the frequency $\nu$ ranging from 8 GHz (upper curve) to 11 GHz (lower curve). The dashed line represents the time $t_l$ of a wave traveling at the light speed, as in Fig. 4a.

In general, the integration in Eq. [19] has to be made numerically. Alternatively, by applying the Jordan Lemma, the integration can be trasformed into a series which may be summed numerically or, in a few special cases, depending on the expression of $A(\omega)$, in a closed form. Analytical details about some special cases are reported in the Appendix.

Figure 6*a*, *b* refers to a δ-function like pulse (case *a* in the Appendix) for three values of the gap *d*. Figures 7*a*, *b* and 8*a*, *b* show the modulus of the transmitted field in two cases relative to rectangular and trapezoidal pulses, respectively (cases *c* and *d* in the Appendix). In all cases, it appears from the figures that, for small values of γ, that is for large Γ (Figs. 6*a*, 7*a*, 8*a*), the trasmitted pulse is practically symmetrical with respect to *t*=0; for the large value of γ (Figs. 6*b*, 7*b*, 8*b*), the trasmitted pulse is strongly asymmetrical.

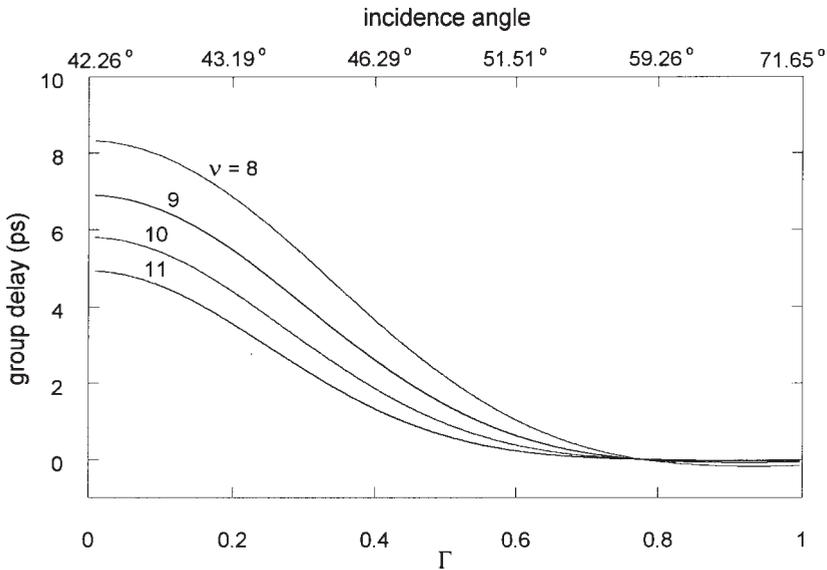

FIG. 5

Group delay, as deduced from Eq. (17), for the same values of the frequency ν as in Fig. 4*c*.

## 3. Conclusions

Here, we have presented a theoretical analysis for describing the propagation of waves and pulses through a forbidden region where evanescent waves take place. Elsewhere (15), we reported on the results of a two prism experiment in which superluminal behaviors, both in the phase and group velocities, are evidenced in the presence of evanescent microwaves. Those results, analogous to those already obtained in previous works (14, 16), can be interpreted on the basis of the analysis developed here. A similar theoretical investigation was also made in Ref. (17), in which superluminal motions of electromagnetic wave packets were predicted. The extent to which such conclusions can hold for the signal velocity (even if limited to small distances, of the order of the wavelength) continues to be a debated and unresolved question also (and mainly) because of the difficulty in giving a unequivocal definition of this quantity. A crucial point could be played by the spectral limitations (always present in the experiments) of the adopted pulse. The theoretical model considered here has not this kind of limitation, but in the numerical computations the finiteness of the spectral extension is imposed by the computational technique. Therefore, only in the analytical results obtained for a pulse like a $\delta$-function this spectral limitation is actually absent and the result obtained appears to be the closest one to the case of a true signal but only from the spectrum point of view. We cannot forget that the model considered here suffers due to the schematizaiton of the problem (infinitely extended boundaries and wavefronts).

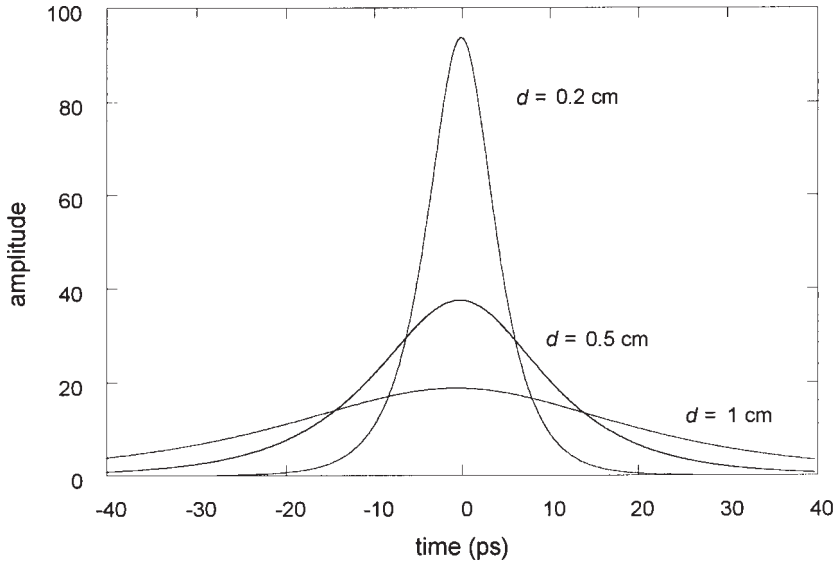

Fig. 6a

Transmitted field, as given by Eq. [A16], for an incident pulse like a Dirac $\delta$-function as a function of time, for different gap width, for $n=1.49$ and $\Gamma=0.8$ ($\gamma=0.51$)

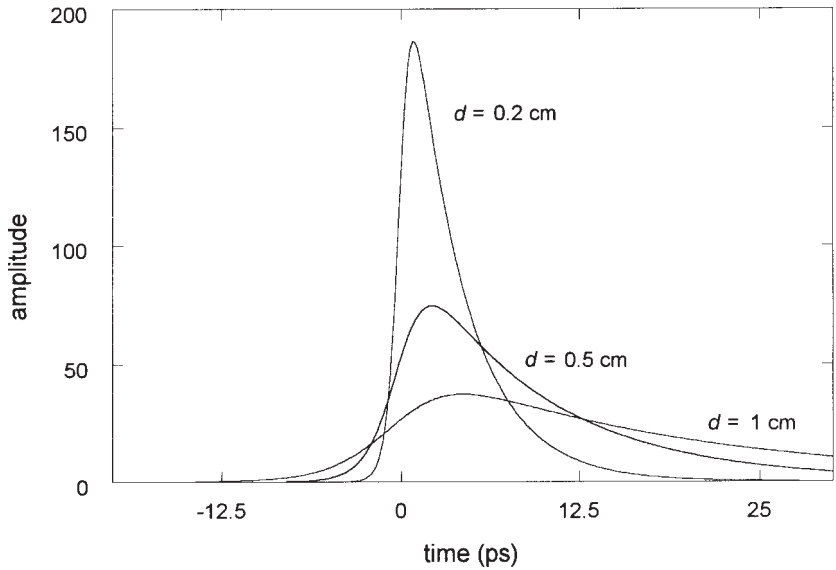

FIG. 6b

Same as Fig. 6a for $\Gamma=0.2$ ($\gamma=0.73$).

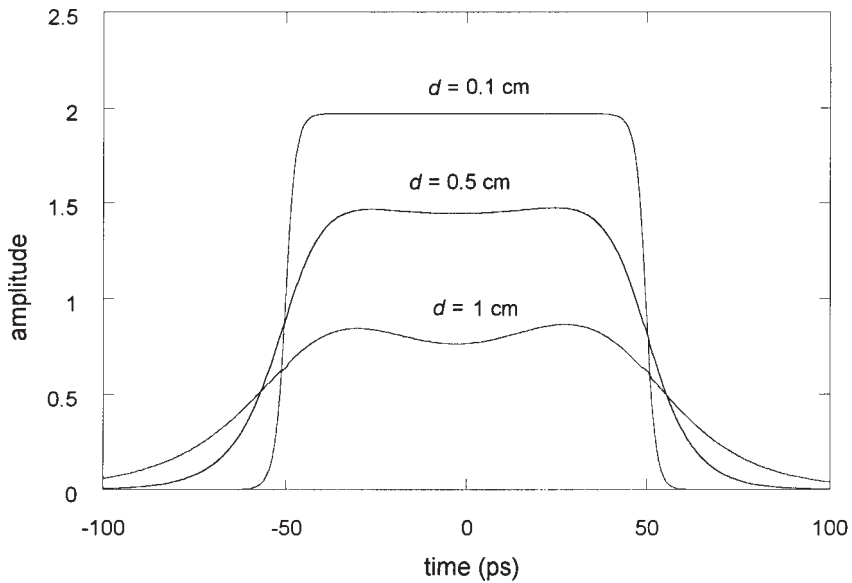

FIG. 7a

Transmitted field for a rectangular incident pulse over a carrier of 10 GHz, as given by Eq. [A25], duration 100 ps, $n=1.49$, $\Gamma=0.8$ ($\gamma=0.51$). The finite rise (and fall) time is due to the spectral limitation in the computer calculations and, increasing the gap, originates the Gibbs effect.

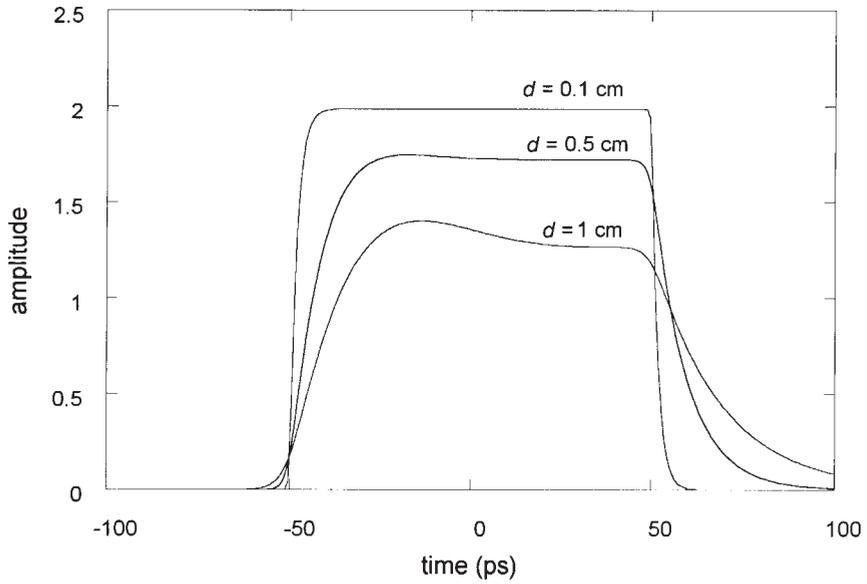

FIG. 7b

Same as Fig. 7a for $\Gamma=0.2$ ($\gamma=0.73$).

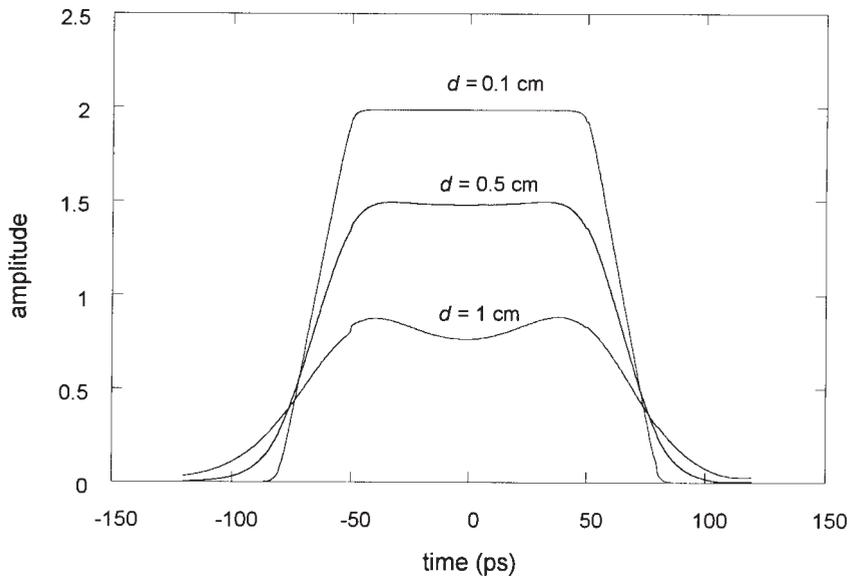

FIG. 8a

The same as in Fig. 7 for a trapezoidal incident pulse Eq. [A31] whose duration is 100 ps, rise and fall times are 30 ps, $\Gamma=0.8$ ($\gamma=0.51$).

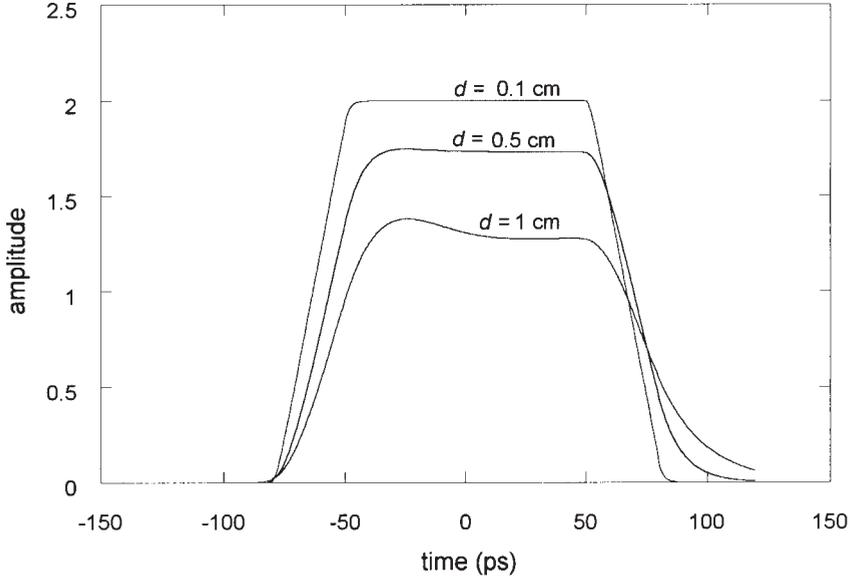

FIG. 8b

Same as Fig. 8a for $\Gamma=0.2$ ($\gamma=0.73$).

## Appendix

We now illustrate how to apply Jordan Lemma in some cases of interest. We have to evaluate integrals of the type

[A1] $\quad I(t, T_0, m) = \exp(i\omega_0 T_0) \int_{-\infty}^{\infty} G^m(\omega-\omega_0)\ \tau(\omega) \exp[-i\omega(t+T_0)] d\omega$

with

[A2] $\qquad\qquad G(\omega-\omega_0) = \dfrac{i}{\omega-\omega_0}$

[A3] $\qquad\qquad \tau(\omega) = iK \dfrac{1}{e^{a\omega+2i\phi} - e^{-a\omega-2i\phi}},$

where

[A4] $\qquad\qquad a = \dfrac{\Gamma d}{c}, \quad \tan\phi = \dfrac{\Gamma}{n\gamma}, \quad K = \dfrac{4n\gamma\Gamma}{(n\gamma)^2 + \Gamma^2}.$

According to Jordan Lemma, we have to distinguish two cases, namely the case in which $t+T_0<0$ and the case in which $t+T_0>0$.

For $t+T_0<0$, the integration path $-\infty, +\infty$ may be closed with a line at infinity in the Im $\omega>0$ half-plane. Hence $I(t, T_0, m)$ may be expressed with the sum of residues of the integrand function at the poles in the upper half-plane, plus (one half of) a possible residue due to the function $G(\omega-\omega_0)$ on the real axis at $\omega=\omega_0$.

For $t+T_0>0$, $I(t, T_0, m)$ may be expressed with the sum of residues (with the sign changed) of the integrand function at the poles in the lower half-plane, minus one half of a possible residue of $G(\omega-\omega_0)$ on the real axis at $\omega=\omega_0$.

The existence of a residue at $\omega=\omega_0$ depends on $m$. For $m=0$, there is no pole and no residue at $\omega=\omega_0$:

[A5] $$R(\omega_0,0)=0.$$

For $m=1$, there is a first-order pole and the corresponding residue $R(\omega_0,1)$ is given by

[A6] $$R(\omega_0,1)=-2\pi\ \tau(\omega_0)\ \exp(-i\omega_0 t).$$

For $m>1$, $I(t, T_0, m)$ diverges too rapidly at $\omega=\omega_0$. However, the difference between two such integrals, corresponding to different values of the parameter $T_0$, may compensate for the $m$-th order divergence and present a first order pole.

As to the poles in the complex $\omega$-plane, these are located at

[A7] $$\omega_j = i\ \Omega_j,$$

where

[A8] $$\Omega_j = \frac{1}{a}(j\pi - 2\phi)$$

hence, they are in the upper half-plane for $j>0$ and in the lower half-plane for $j\leq 0$. At these poles, the residues are given by

[A9] $$R(\Omega_j) = F(m, j, T_0)(-1)^j \exp[\Omega_j(t+T_0)], \quad m=0,1$$

where

$$F(m, j, T_0) = K'M_0^m \exp\left[i(m\psi + \omega_0 T_0)\right]$$

[A10] $$K' = -\frac{\pi K}{a}, \quad M_0 = \frac{1}{\sqrt{\Omega_j^2 + \omega_0^2}}$$

$$\cos\psi = \frac{\Omega_j}{M_0}, \quad \sin\psi = -\frac{\omega_0}{M_0}.$$

We conclude that for $m=0,1$ and $t+T_0<0$,

[A11] $$I(t,T_0,m) = I^-(t,T_0,m) = \frac{1}{2}R(\omega_0,m) + \sum_{j>0} F(m,j,T_0)(-1)^j \exp\left[\Omega_j(t+T_0)\right]$$

whereas, for $t+T_0>0$,

[A12] $$I(t,T_0,m) = I^+(t,T_0,m) = -\frac{1}{2}R(\omega_0,m) - \sum_{j\leq 0} F(m,j,T_0)(-1)^j \exp\left[\Omega_j(t+T_0)\right].$$

Let us now apply the above expressions to some particular cases.

**Case a** - *The δ(t) pulse*

The spectrum of the δ-function is a constant

[A13] $\qquad\qquad\qquad A(\omega)=1.$

Accordingly Eq. [19] may be written as

[A14] $\qquad E^t = \dfrac{E_0}{2\pi} \int_{-\infty}^{\infty} \tau(\omega)\exp(-i\omega t)d\omega = \dfrac{E_0}{2\pi} I(t,0,0)$

which is of the same type as Eq. [A1] with $T_0=0$, $m=0$, and therefore, by using Eqs. [A5], [A11] and [A12], and considering that $F(0, j, 0)=K'$, we obtain

[A15]
$$E^t = \dfrac{E_0 K'}{2\pi} \sum_{j>0} (-1)^j \exp(\Omega_j t), \quad \text{for } t<0$$
$$E^t = -\dfrac{E_0 K'}{2\pi} \sum_{j\leq 0} (-1)^j \exp(\Omega_j t), \quad \text{for } t>0.$$

Taking into account Eq. [A8], the two series of Eqs. [A15] are geometrical series which can be summed up, giving the same result both for $t<0$ and $t>0$,

[A16] $\qquad E^t = -\dfrac{E_0 K'}{2\pi} \exp(-2\phi ct/\Gamma d) \left[1+\exp(-\pi ct/\Gamma d)\right]^{-1}.$

It may be proved, by a direct integration of Eq. [19], that Eq. [A16] also holds for $t=0$.

**Case b** - *The step-function over a carrier at frequency $\omega_0$.*

The spectrum of a positive step function, of amplitude $A_0$ located at $t=0$ and modulating a carrier at frequency $\omega_0$, is given by

[A17] $\qquad\qquad A(\omega) = A_0 \left[ \pi\delta(\omega-\omega_0) + \dfrac{i}{\omega-\omega_0} \right].$

Introducing Eq. [A17] into Eq. [19] yields

[A18] $\quad E^t = \dfrac{1}{2} E_0 A_0 \tau(\omega_0) \exp(-i\omega_0 t) + \dfrac{1}{2\pi} E_0 A_0 \int_{-\infty}^{\infty} \dfrac{i}{\omega-\omega_0} \tau(\omega)\exp(-i\omega t)d\omega.$

The integral

$$J = \int_{-\infty}^{\infty} \dfrac{i}{\omega-\omega_0} \tau(\omega) \exp(-i\omega t) d\omega$$

is of the same type as in Eq. [A1], with $T_0=0$, $m=1$. Therefore, taking into account Eqs. [A6], [A8], [A11] and [A12], we can write,

[A19] $$J = \frac{1}{2}R(\omega_0,1) + \sum_{j>0} F(1,j,0)(-1)^j \exp(\Omega_j t), \quad \text{for } t<0$$

[A20] $$J = -\frac{1}{2}R(\omega_0,1) - \sum_{j\leq 0} F(1,j,0)(-1)^j \exp(\Omega_j t), \quad \text{for } t>0$$

By introducing Eqs. [A19] and [A20] into Eq. [A18], and considering that $F(1, j, 0) = K'M_0 \exp(i\psi)$, we can conclude that

[A21]
$$E^t = \frac{E_0 A_0}{2\pi} K' \sum_{j>0} (-1)^j M_0 \exp(i\psi) \exp(\Omega_j t), \quad \text{for } t<0$$

$$E^t = E_0 A_0 \left[ \tau(\omega_0) \exp(-i\omega_0 t) - \frac{K'}{2\pi} \sum_{j\leq 0} (-1)^j M_0 \exp(i\psi) \exp(\Omega_j t) \right], \quad \text{for } t>0$$

**Case c** - *A rectangular pulse carried by a frequency $\omega_0$*

For a rectangular pulse of height $A_0$ and duration from $-T$ to $T$, the spectrum $A(\omega)$ may be written as

[A22] $$A(\omega) = -i\frac{A_0}{\omega - \omega_0}\left[\exp\left[i(\omega - \omega_0)T\right] - \exp\left[-i(\omega - \omega_0)T\right]\right].$$

By introducing Eq. [A22] into Eq. [19], we can write

[A23] $$E^t = -\frac{1}{2\pi} E_0 A_0 [J_1 - J_2],$$

where $J_1$ and $J_2$ are of the same type as Eq. (A1), that is,

$$J_1 = I(t,-T, 1), \quad J_2 = I(t, T, 1)$$

Therefore, by applying Eqs. (A11) and (A12) we can write

[A24]
$$J_1 = I^-(t,-T,1), \quad \text{for } t-T<0, \ t<T$$
$$J_1 = I^+(t,-T,1), \quad \text{for } t-T>0, \ t>T$$
$$J_2 = I^-(t,T,1), \quad \text{for } t+T<0, \ t<-T$$
$$J_2 = I^+(t,T,1), \quad \text{for } t+T>0, \ t>-T$$

and, by introducing into Eq. [A23],

[A25]
$$E^t = -\frac{E_0 A_0}{2\pi}\left[I^-(t,-T,1) - I^-(t,T,1)\right], \quad \text{for } t<-T$$

$$E^t = -\frac{E_0 A_0}{2\pi}\left[I^-(t,-T,1) - I^+(t,T,1)\right], \quad \text{for } -T<t<T$$

$$E^t = -\frac{E_0 A_0}{2\pi}\left[I^+(t,-T,1) - I^+(t,T,1)\right], \quad \text{for } t>T$$

Note that the two terms $R(\omega_0, m)$ appearing in the $I$ integrals cancel one another for $t<-T$ and $t>T$, whereas they sum in the interval $-T<t<T$.

**Case d** - *A trapezoidal pulse carried by a frequency $\omega_0$.*

In this case the spectrum $A(\omega)$ of the pulse is given by

[A26]
$$A(\omega) = -\frac{A_0}{D}\frac{1}{(\omega-\omega_0)^2}\{\exp[-i(\omega-\omega_0)T_1] - \exp[-i(\omega-\omega_0)T] \\ -\exp[i(\omega-\omega_0)T] + \exp[i(\omega-\omega_0)T_1]\},$$

where the parameters are indicated in Fig. 8. The four exponentials in the right hand side of Eq. [A26], introduced into Eq. [19], cannot be handled separately, because of the second-order singularity at $\omega=\omega_0$. However we can proceed as follows.

Upon introduction of Eq. [A26], Eq. [19] can be written as

[A27]
$$E^t = \frac{E_0}{2\pi}\exp(-i\omega_0 t)L(t),$$

where

[A28]
$$L(t) = \int_{-\infty}^{\infty} A(\omega)\tau(\omega)\exp[-i(\omega-\omega_0)t]\,d\omega.$$

It follows that

[A29]
$$\frac{dL}{dt} = \int_{-\infty}^{\infty} A(\omega)\left[-i(\omega-\omega_0)\right]\tau(\omega)\exp[-i(\omega-\omega_0)t]\,d\omega$$

which, after the introduction of Eq. [A26], can be written as

$$\frac{dL}{dt} = \frac{A_0}{D}e^{i\omega_0 t}\int_{-\infty}^{\infty}\frac{i}{\omega-\omega_0}\{\exp[-i(\omega-\omega_0)T_1] - \exp[-i(\omega-\omega_0)T] \\ -\exp[i(\omega-\omega_0)T] + \exp[i(\omega-\omega_0)T_1]\}\tau(\omega)e^{-i\omega t}d\omega$$

that is,

[A30]
$$\frac{dL}{dt} = \frac{A_0}{D}\exp(i\omega_0 t)\left[I(t,T_1,1) - I(t,T,1) - I(t,-T,1) + I(t,-T_1,1)\right],$$

where function $I$ is defined by Eq. [A1]. By using Eqs. [A11] and [A12], we have

$$\frac{dL}{dt} = \frac{A_0}{D}e^{i\omega_0 t}\left[I^-(t,T_1,1) - I^-(t,T,1) - I^-(t,-T,1) + I^-(t,-T_1,1)\right], \text{ for } t<-T_1$$

$$\frac{dL}{dt} = \frac{A_0}{D}e^{i\omega_0 t}\left[I^+(t,T_1,1) - I^-(t,T,1) - I^-(t,-T,1) + I^-(t,-T_1,1)\right], \text{ for } -T_1<t<-T$$

$$\frac{dL}{dt} = \frac{A_0}{D}e^{i\omega_0 t}\left[I^+(t,T_1,1) - I^+(t,T,1) - I^-(t,-T,1) + I^-(t,-T_1,1)\right], \text{ for } -T<t<T$$

$$\frac{dL}{dt} = \frac{A_0}{D}e^{i\omega_0 t}\left[I^+(t,T_1,1) - I^+(t,T,1) - I^+(t,-T,1) + I^-(t,-T_1,1)\right], \text{ for } T<t<T_1$$

$$\frac{dL}{dt} = \frac{A_0}{D}e^{i\omega_0 t}\left[I^+(t,T_1,1) - I^+(t,T,1) - I^+(t,-T,1) + I^+(t,-T_1,1)\right], \text{ for } t>T_1$$

To obtain L(ω), which when introduced into Eq. [A27] yields $E^t$, we have to integrate the above expressions with respect to time. By recalling Eqs. [A11] and [A12], we can write

$$N^{\pm}(t, T_0) = \int \exp(i\omega_0 t) I^{\pm}(t, T_0, 1) dt$$

$$= \pm \pi \ \tau(\omega_0) t \mp \exp(i\omega_0 t) \sum_j F(2, j, T_0)(-1)^j \exp\left[\Omega_j(t + T_0)\right] + \text{const.}$$

where $F(2, j, T_0) = F(1, j, T_0)/(\Omega + i\omega_0)$ and the sum is limited to $j>0$ for the upper sign, and to $j \leq 0$ for the lower sign.

Note that the contributions of the residues cancel out in the time intervals $t<-T_1$, $-T<t<T$, $t>T_1$.

As the integration constants, we are interested in those of $L(t)$ or $E^t$, not in those of $N^{\pm}(t, T_0)$ separately. Accordingly, they can be deduced by considering that

$$L(-\infty) = 0$$

and that $L(t)$ is continuous at $-T_1$, $-T$, $T$, $T_1$. Thus, for example,

$$L(-T_1+0) = L(-T_1-0).$$

In conclusion, we can write

[A31]
$$E^t = \frac{E_0 A_0}{2\pi D} e^{-i\omega_0 t} \left[N^-(t, T_1) - N^-(t, T) - N^-(t, -T) + N^-(t, -T_1)\right] +$$
$$+ \text{const, for } t < -T_1$$

$$E^t = \frac{E_0 A_0}{2\pi D} e^{-i\omega_0 t} \left[N^+(t, T_1) - N^-(t, T) - N^-(t, -T) + N^-(t, -T_1)\right] +$$
$$+ \text{const, for } -T_1 < t < -T$$

$$E^t = \frac{E_0 A_0}{2\pi D} e^{-i\omega_0 t} \left[N^+(t, T_1) - N^+(t, T) - N^-(t, -T) + N^-(t, -T_1)\right] +$$
$$+ \text{const, for } -T < t < T$$

$$E^t = \frac{E_0 A_0}{2\pi D} e^{-i\omega_0 t} \left[N^+(t, T_1) - N^+(t, T) - N^+(t, -T) + N^-(t, -T_1)\right] +$$
$$+ \text{const, for } T < t < T_1$$

$$E^t = \frac{E_0 A_0}{2\pi D} e^{-i\omega_0 t} \left[N^+(t, T_1) - N^+(t, T) - N^+(t, -T) + N^+(t, -T_1)\right] +$$
$$+ \text{const, for } t > T_1$$